\documentclass[twocolumn,showpacs,preprintnumbers,amsmath,amssymb]{revtex4}
\usepackage{graphicx}
\usepackage{dcolumn}
\usepackage{bm}
\usepackage{amsmath}
\usepackage{amsfonts}%

\begin{document}

\title{Formation of the simplest stable negative molecular ion H$_3^-$ in interstellar medium}
\author{}
\author{V. Kokoouline$^{1,2}$, M. Ayouz$^{1}$, R. Gu\'erout$^{1}$, M. Raoult$^{1}$, J. Robert$^{1}$, and O. Dulieu$^{1}$}
\affiliation{$^{1}$Laboratoire Aim\'e Cotton, CNRS, B\^at 505, Universit\'e Paris 11, 91405 Orsay Cedex, France\\$^{2}$Department of Physics, University of Central Florida, Orlando, Florida 32816, USA }


\begin{abstract} 
We present the theory of radiative association of atoms and molecules, and we apply it to the (H$_2$-H$^-$) van der Waals complex. We discuss the possibility for  the H$_3^-$ ion to be formed in the interstellar medium in an environment with abundant ionized molecular hydrogen H$_2^+$. The observation of H$_3^-$ would also be a probe for the presence of H$^-$  in the interstellar medium. By computing the electronic structure of the H$_3^-$ ion, we determine its dipole moment, bound states, rotational constants,  predissociated vibrational resonances and their lifetimes, and suggest a way to detect the ion in the interstellar medium. 
\end{abstract}

\pacs{98.38.Dq, 95.30.Ft, 33.80.-b}

\maketitle

Many chemical reactions in the interstellar medium (ISM) are powered by cosmic rays: Atoms and molecules (mainly molecular hydrogen) are ionized by the radiation that provides sufficient energy to initiate a chain of chemical reactions in interstellar clouds leading to the synthesis of polyatomic molecules. A number of positive ions have been observed and identified in the ISM, in particular, the H$_3^+$ ion. It is the simplest triatomic positive ion that plays an important role in chemistry and evolution of interstellar clouds \cite{geballe96,oka06b}, as its abundance is strongly related to the production of H$_2^+$ in the ISM.  In contrast, only a few negative ions have been detected so far in the ISM: C$_3$N$^-$,  C$_4$H$^-$, C$_6$H$^-$, and C$_8$H$^-$ \cite{thaddeus08}. While quite stable, the simplest negative triatomic ion, H$_3^-$ (predicted to be bound by about 0.013 eV \cite{starck93}) has not been detected so far in the ISM. In this article we argue that the H$_3^-$ ion is indeed formed in cold (below 150 K) interstellar clouds, provided that H$_2^+$ (and free electrons) are available. Using a recently published potential surface \cite{panda04}, we determine its main spectroscopic properties and suggest possible ways to observe it in the ISM. We analyze the possible formation mechanism of H$_3^-$ in the ISM in collisions between H$^-$ and H$_2$.
In addition as H$^-$ has only one bound state and, therefore cannot be directly observed, the detection of H$_3^-$ would be a probe for the presence of H$^-$ in the ISM, which is believed to exist in the ISM but has not been detected so far.

The chemistry of interstellar clouds is initiated by ionization of molecular hydrogen by cosmic rays with a typical rate constant  $\zeta\sim3\times  10^{-17}$s$^{-1}$ in diffuse interstellar clouds \cite{oka06b}.  (Cloud densities are $\sim 10^2$ cm$^{-3}$ in diffuse and  $\sim 10^4$ cm$^{-3}$ in dense clouds.) The ionized molecular hydrogen H$_2^+$ quickly forms H$_3^+$ in collisions with H$_2$, with a rate constant  $\sim 2\times 10^{-9}$cm$^3$/s \cite{oka06b}. The escaped electron has a large kinetic energy and undergos many elastic collisions with environmental H$_2$ before its rethermalization. Possible inelastic $e^-$+H$_2$ collisions will lead to vibrational excitation of H$_2$, and to dissociative attachment (DA) $e^-$+H$_2\to$ H+H$^-$ for collision energies above the threshold at 3.7 eV, with a cross section of about $\sigma_{DA}\sim 10^{-21}$~cm$^2$ \cite{schulz65,horacek04}. Therefore, because the DA reaction rate per one H$_2$ molecule  is larger than the rate of electron production $\zeta$,  $v_e\sigma_{DA}n(\mathrm{H}_2)>\zeta$, the H$^-$ ions are produced in the ISM with a (binary) rate constant mainly determined by the product $\zeta n(\mathrm{H}_2)$. Note that the same rate constant also determines the production rate of H$_3^+$ in the ISM \cite{oka06b}. 


While the collisions between H$^-$ and H$_2$ molecules have been studied both theoretically \cite{starck93,panda04} and experimentally \cite{muller96,wester09}, the structure of the H$_3^-$ ion has been rarely studied in the past \cite{starck93}, and a single indirect observation has been reported in laboratory plasmas \cite{wang03}. The H$_3^-$ ion is well represented as a van der Waals complex (H$_2 \cdots $H$^-$) \cite{starck93}. The molecule has several rovibrational states, bound by about 20-100~cm$^{-1}$ for the lowest ones (see Table \ref{tab:bound_states}). According to the present study, there are also a number of predissociated resonances, which can be described as excited rovibrational states $(j,v_d)$ of H$_2$ perturbed by H$^-$, coupled to the dissociation continuum H$_2(j',v_d')$+H$^-$ with energy of the dimer state $E(j',v_d')$ lower than $E(j,v_d)$. We found that the widths of the broadest resonances in the low energy spectrum (below 4000~cm$^{-1}$) are in the range of $0.2-1.5$~cm$^{-1}$, which corresponds to lifetimes of $3.5-26$~ps.

To form a bound H$_3^-$ molecule in H$_2$+H$^-$ collisions in the ISM two mechanisms are possible: three-body recombination (TBR) or radiative association (RA):
\begin{eqnarray}
 \mathrm{H}_2+\mathrm{H}^-+\mathrm{X} \to  \mathrm{H}_3^- +\mathrm{X}\,\  :\ \mathrm{TBR}\,,\\
 \mathrm{H}_2+\mathrm{H}^- \to  \mathrm{H}_3^- +\hbar\omega\,\  :\ \mathrm{RA}\/.
\end{eqnarray}
The decay of H$_3^-$ in diffuse clouds is determined by collisions with H$_3^+$, other positive ions, and by interstellar radiation.  The TBR rate constant $k_{3b}$ can roughly be estimated considering just geometrical (van der Waals) sizes of reactants and procedure outlined in Ref. \cite{glosik09a}. We obtained $k_{3b}\sim 10^{-27}-10^{-30}$ cm$^{6}$/s. With typical number densities $n=10^2-10^4$ cm$^{-3}$ in the ISM, the three-body recombination as a way to form H$_3^-$ is expected to be much slower than possible two-body processes involving H$^-$. Using the theory developed below, we estimated the RA rate coefficient $k_{\rm RA}$ to be about $4\times 10^{-21}$ cm$^3/$s, corresponding to a survival lifetime of H$^-$ ions $\tau_{\rm RA}=1/(k_{\rm RA}n)$ with respect to RA. The lifetime should be compared with lifetimes with respect to other processes that remove H$^-$ from the interstellar gas: photodetachment of an electron from H$^-$ and recombination with positive ions. We have estimated the lifetime with respect to photodetachment using the available data about the photoabsorption spectrum of H$^-$ \cite{rau96} and intensity of radiation in typical diffuse interstellar clouds \cite{oka06b} and obtained the upper limit for the lifetime $\tau_{PD}\sim 10^{16}$ s. We have also estimated the lifetime $\tau=1/(n_+ k^+)$ with respect to recombination of H$^-$ with positive molecular ions in the ISM, in particular, with H$_3^+$ ($n_+\sim 10^{-6}$cm$^{-3}$, the corresponding rate constant $k^+\sim 10^{-9}$cm$^3/$s is estimated using geometrical sizes of H$_3^+$ and H$^-$).   Therefore, we predict that the  H$_3^-$ ion is produced in cold (especially dense) interstellar clouds consisted of molecular hydrogen. We also point out that if the assumption that an important fraction of all electrons produced by ionization of H$_2$ would form H$^-$ is correct, the destruction of H$_3^+$ in diffuse clouds would be then determined not only by the dissociative recombination but also the recombination with H$^-$. Such a hypothesis would solve the enigma of the anomalously large column density of H$_3^+$ in diffuse clouds \cite{oka06b}.

\begin{table}[tbp]
\begin{tabular}{|p{2.0cm}|p{2.2cm}|p{1.4cm}|p{1.5cm}|}
\hline
 $J,j,\Omega,v_t,v_d,\Gamma$&  Energies, cm$^{-1}$&$B_Z$, cm$^{-1}$&$B_X$, cm$^{-1}$\\
\hline
 $ 0,0,0,0,0,A_{1}' $  &-105.01&211.2&3.23\\
 $ 0,0,0,0,0,E' $ &-104.8&211.&3.\\
 $ 0,1,0,0,0,E'$ &-44.0&345.&3.7\\
$ 0,1,0,0,0,A_{2}'$ &-43.90&345.1&3.67\\
$ 0,0,0,1,0,A_{1}'$ &-17.86&187.5&1.8\\
 $ 0,0,0,1,0,E' $ &-17.7&188.&2.\\
 $ 0,1,0,1,0,A_{2}'$&73.24&311.9&2.36\\
\hline
\end{tabular}
\caption{Computed binding energies (with respect to the lowest dissociation limit H$_2(0, 0)+$H$^-$) and rotational constants of bound rovibrational states of H$_3^-$ relative to the $Z$ and $X$ axes. The quantum numbers are defined in the text.}
\label{tab:bound_states}
\end{table}

\begin{figure}
\includegraphics[width=8.cm]{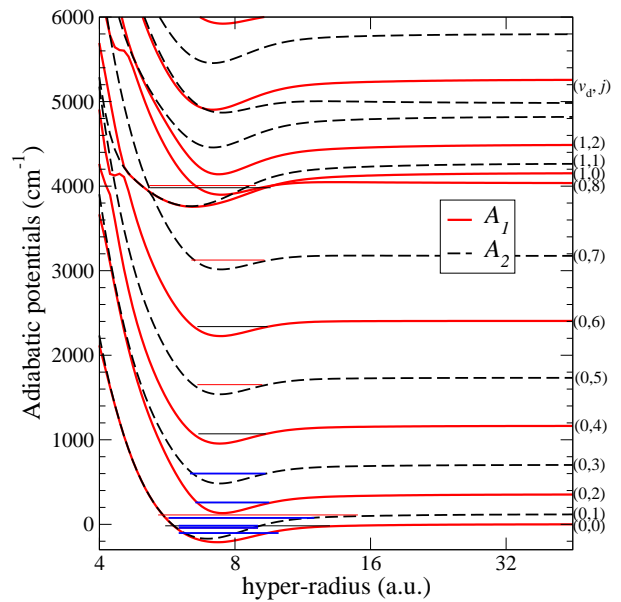}
\caption{(Color online)  H$_3^-$ hyperspherical adiabatic curves for $J=0$. Only $A_1$ and $A_2$ vibrational symmetries are shown. The curve of the $E$ symmetry are very similar to the  $A_1$ and  $A_2$ curves for energies below 4000 cm$^{-1}$ (due to the barrier for the proton exchange \cite{starck93}), and are not displayed here for clarity. Horizontal lines indicate positions of several bound states (reported in Table \ref{tab:bound_states}) and resonances. (Here and below, a.u. means atomic units)}
\label{fig:ad_curves}
\end{figure}

We have performed the numerical calculation of  bound states, their rotational constants, predissociated resonances, and cross-sections within the Complete Nuclear Permutation Inversion (CNPI) group \cite{bunker98} $D_{3h}$ of the molecule, using the H$_3^-$ potential surface from Ref. \cite{panda04}. The details of the numerical procedure are given in Refs. \cite{kokoouline06,blandon07}.  We solved numerically the three-dimensional Schr\"odinger equation for the molecule in hyperspherical coordinates, separating hyperangles from the hyper-radius and using the slow variable discretization.
 Figure \ref{fig:ad_curves} yields an insight into the H$_3^-$ hyperspherical adiabatic curves calculated for $J=0$. Each curve at large hyper-radii is correlated with a H$_2(v_d,j)+$H$^-$ dissociation limit. The lowest bound states and resonances can be characterized by the approximate quantum numbers $j,\Omega,v_t,v_d$ defined below.

{\it Symmetry and approximate quantum numbers for ${\mathrm H}_3^-$.}
The H$_3^-$  molecule consists of atoms with three identical nuclei, described within the CNPI group $D_{3h}$. However, due to the van der Waals nature of the H$_3^-$ ion in its low energy states \cite{starck93}, the study of the lowest rovibrational states of H$_3^-$ can be performed within the Molecular Symmetry group (or MS group \cite{bunker98} $C_{2v}$, which is a subgroup of the CNPI group above).  Therefore, in order to identify allowed H$_3^-$ states for ortho- and para-configurations of the total nuclear spin $I$ ($I=3/2$ and 1/2, respectively) we derive below approximate wave functions and quantum numbers of the lowest rovibrational states. 

The H$_3^-$ molecule has a definite total angular momentum $J$ and its projection $m$ on the $z$-axis of the space-fixed coordinate system (SCS). The orientation of the molecular coordinate system (MCS) with respect to the SCS is given by the Euler angles $\alpha$, $\beta$ and $\gamma$. The MCS $Z$-axis connects the center of the H$_2$ dimer with the nuclei of H$^-$, and its $X$-axis is in the molecular plane. In the present model, we assume that the H$_2$ dimer is characterized by the vibrational quantum number $v_d$ and its angular momentum $j$ with projection $\Omega$ on the $Z$ axis. Its orientation with respect to $Z$ is given by the azimuthal angle $\theta$. Assuming that the vibrational and rotational motions are uncoupled, the H$_3^-$ wave function is represented by (omitting a normalization factor for simplicity):
\begin{equation}
\label{eq:wf}
|J,j,\Omega,v_t,v_d\rangle = \left[ D_{m\Omega}^{J}(\alpha,\beta,\gamma)\right]^*P_{j}^{\Omega}(\cos\theta)|v_d\rangle |v_t\rangle\,,
\end{equation}
where the associated Legendre polynomial $P_{j}^{\Omega}(\theta)$ describes the H$_2$ rotational state, $D_{m\Omega}^{J}(\alpha,\beta,\gamma)$ is a Wigner function, and $v_t$ is a quantum of motion along the $Z$ axis. The H$_3^-$ molecule is then considered as a symmetric rotor, with two of the three moments of inertia (along the $X$ and $Y$ axis) almost equal to each other.  $m$ does not influence the energy of the state, and will be omitted. The energy of the state $E(|J,j,\Omega,v_t,v_d\rangle)$ is given (approximately) by:
\begin{eqnarray}
E(|J,j,\Omega,v_t,v_d\rangle)=E_{v_d}+E_{v_t}+B_Z j(j+1) + \nonumber\\
\left[B_X J(J+1)+(B_Z-B_X)\Omega^2\right]\,.
\end{eqnarray}
where $B_X$ and $B_Z$ are the rotational constants with respect to the $X$- and $Z$-axes, respectively. The wave functions  $|J,j,\Omega,v_t,v_d\rangle$ transform under $C_{2v}$ operators $(12)$ (permutation of nuclei of the H$_2$ dimer) and $E^*$ (inversion \cite{bunker98}) as:
\begin{eqnarray}
\label{eq:12E}
(12)|J,j,\Omega,v_t,v_d\rangle=(-1)^j|J,j,\Omega,v_t,v_d\rangle\nonumber\,\\
E^*|J,j,\Omega,v_t,v_d\rangle=(-1)^J|J,j,-\Omega,v_t,v_d\rangle\,.
\end{eqnarray}
where we assumed that H$_2$ is in the $X^1\Sigma_g^+$ ground state and the trimer in the $^1A_1'$ state.
The final step is to ensure that the approximate wave functions of H$_3^-$ have a proper symmetrization with respect to exchange of identical nuclei, using the projection operators $\hat P_\Gamma$ on a particular irreducible representation (irrep.) $\Gamma$ of the CNPI group $D_{3h}$. The symmetrized states will be referred to as  $|J,j,\Omega,v_t,v_d,\Gamma\rangle$.
For certain combination of quantum numbers $J$, $j$ and $\Omega$, some of the projections $\hat P_\Gamma$ are zero. It means that the corresponding irrep. $\Gamma$ is not allowed for this set of quantum numbers \cite{douguet08a}. The general rules are derived from Eqs. (\ref{eq:12E}): both (even and odd, labelled as prime and double prime)  parities are allowed for non-zero $\Omega$. If $\Omega=0$, the parity is given by $(-1)^J$. The $E$ irrep. is allowed for any combination of $J,j,\Omega$ (assuming that the parity $E'$ or $E''$ is given by the above rule). The $A_1'$ and $A_1''$ (resp. $A_2'$ and $A_2''$) irreps. are allowed for even (resp. odd) $j$.

With such wave functions, we can determine the allowed states for para-  and ortho-H$_3^-$. Because the three nuclei are identical fermions, the total wave function (including the nuclear spin factor) of H$_3^-$ can only be of $A_2'$ or $A_2''$ irrep. Since the para-H$_3^-$ nuclear spin part of the wave function transforms as $E'$ in $D_{3h}$, the allowed spatial irrep. could be $E'$ or $E''$. For ortho-H$_3^-$ ($A_1'$ irrep. of the nuclear spin), only $A_2'$ and $A_2''$ rovibrational wave functions are allowed. The lowest rovibrational state $|0,0,0,0,0,A_1'\rangle$ is not allowed for H$_3^-$ (but allowed for D$_3^-$). The lowest allowed state is $|0,0,0,0,0,E'\rangle$ is the para-H$_3^-$ state. The lowest $A_2'$ rotational state (lowest ortho-H$_3^-$) is $|0,1,0,0,0,A_2'\rangle$.

{\it Theory of radiative association in dimer-atom collisions.}
In order to estimate the cross-section and the rate coefficient for RA, we develop a theoretical framework to treat the radiative association of a dimer and an atom. Our approach is based on theory developed by Herzberg \cite{herzberg50},  and later used by several authors \cite{zydelman90,stancil93,gianturco97} for diatomic molecules, and for photoassociation of cold atoms \cite{cote96,pillet97}. 

In order to adapt the theory to triatomic systems, similarly to Ref. \cite{gianturco97}, we express the Einstein coefficient $A_{q',v_t';q,E}$ for the photon emission from an H$_3^-$ rovibrational state specified by quantum numbers $q=\{J,j,\Omega,v_d,\Gamma\}$ during a H$_2+$H$^-$ collision at an energy  $E$. After a photon of energy $\hbar \omega$ is emitted, the triatomic H$_3^-$ ion is in a state specified by the quantum numbers $q'=\{J',j',\Omega',v_d',\Gamma'\}$ and by $v_t'$ for the vibrational motion of the H$_2 \cdots$H$^-$ van der Waals system. The Einstein coefficient $A_{q',v_t';q,E}$ is given (in a.u.) by:
\begin{eqnarray}
A_{q',v_t';q,E}=\frac{4\omega^3}{3c^3}\lvert \vec r_{q',v_t';q,E}\rvert ^2\,,
\label{eq:einstein}
\end{eqnarray}
where $\lvert \vec r_{q',v_t';q,E}\rvert$ is the matrix element of the dipole moment (with three components $r^{\sigma}$, $\sigma=-1,0,+1$). The value $\lvert \vec r_{q',v_t';q,E}\rvert ^2$ can be evaluated using a technique similar to the one presented in Ref.  \cite{bunker98}: we used Eq. (14-33) of  Ref.  \cite{bunker98} for the line strength and average it over the initial states. In the present model, each rotational state of H$_3^-$ is characterized by a single symmetric top rotational function, so Eq. (\ref{eq:einstein}) reduces to: 
\begin{eqnarray}
A_{q',v_t';q,E}=\frac{4\omega^3}{3c^3}(2J'+1)\times&\\
\times\biggr| \sum_{\sigma}\langle j',\Omega',v_t',v_d' ,\Gamma'\arrowvert \mu^\sigma\arrowvert  j,\Omega,E,v_d,\Gamma\rangle &\left( 
\begin{array}{lll}
 J& 1 & J' \\ 
\Omega & \sigma & -\Omega'
\end{array}
\right)\biggr| ^2\,,\nonumber
\end{eqnarray}
where $\mu^\sigma$ is the $\sigma$ component of the dipole moment calculated in the MCS. The initial collisional state in the above expression is energy normalized. The probability $P_{q',v_t';q,E}$ of an RA event is given by the Einstein coefficient divided with the current density in the flux of incident particles $1/(2\pi)$ for the energy normalized wave function. Finally, the RA cross-section is given by $\pi P_{q',v_t';q,E}/k^2$:
\begin{eqnarray}
\label{eq:cs}
\sigma_{q',v_t';q,E}=\frac{8\pi^2\omega^3}{3k^2c^3}(2J'+1)\times&\\
\times\biggr| \sum_{\sigma}\langle j',\Omega',v_t',v_d' ,\Gamma'\arrowvert \mu^\sigma\arrowvert  j,\Omega,E,v_d,\Gamma\rangle &\left( 
\begin{array}{lll}
 J& 1 & J' \\ 
\Omega & \sigma & -\Omega'
\end{array}
\right)\biggr| ^2\,.\nonumber
\end{eqnarray}
To obtain the cross section $\sigma_{q}(E)$ for the formation of any H$_3^-$ bound state $\{q',v_t'\}$, we have to sum over $q'$ and $v_t'$. Since the nuclear spin is conserved during the RA process, there is no need to include the nuclear spin degeneracy factor: after averaging over the initial state and summing up over final states, the factor will be one. 

The rate constant is finally obtained by a standard integration over a Maxwell-Boltzmann distribution. In the integration, the nuclear spin degeneracy factors $(2I+1)$ as well as the rovibrational energy of the initial state of H$_2$  should be taken into account (see, for example, Ref. \cite{santos07} for details on the averaging procedure). The rate constant could be accurately calculated provided that the dipole moment functions are known over all the configuration space, which is not the case with the available {\it ab-initio} calculations. Here we propose a rough estimation of the rate constant to verify if the RA process is competitive with other processes in the ISM leading to the removal of H$^-$ (if present) from the interstellar gas. For given values of $\Omega$ and $\Omega'$, only one term in the sum of Eq.  (\ref{eq:cs}) is not zero. The $3j$ symbol and the symmetry $\Gamma_\mu=A_1''\oplus A_2''\oplus E''$ of the vector of dipole moment $(\mu^{-1},\mu^0,\mu^{+1})$ in the  $D_{3h}$ group determine selection rules: $J\to J'=J\pm1;\ \Omega\to\Omega'=\Omega,\Omega\pm1$. In addition,  parities of the initial and final states should be opposite.

The largest vibrational dipole moment matrix element in Eq. (\ref{eq:cs}) is expected  when $\Omega=\Omega'$ (because $|\mu^0|$ is much larger than $|\mu^{\pm 1}|$), $j=j'$, and $v_d=v_d'$. When all internal ``vibrational'' quantum numbers ($v_d,\Omega,j$) are the same except $v_t$, the vibrational dipole moment matrix element can be estimated as $\mu^0$ calculated at the equilibrium geometry multiplied with the vibrational Franck-Condon overlap. The Franck-Condon overlap can be approximated using the density of discrete states $1/\Delta E$ calculated at the energy of the final bound state $v_t'$. The value of $3j$ can be taken to be 1 for the allowed transitions ($J\to J\pm1;\ \Omega\to\Omega'=\Omega$) in the rough estimation. The equilibrium value of $\mu^0$ is 4 a.u. \cite{guerout09}, $\Delta E\sim 30-60$ cm$^{-1}$ or $1.5-3\times 10^{-4}$ a.u.; $\hbar\omega\sim 100$ cm$^{-1}$ or $5\times 10^{-4}$ a.u.; $k^2/(2m)\sim 1.3\times 10^{-4}$ a.u (it corresponds to 40K, a reasonable temperature for cold diffuse clouds), $m\sim1200$ a.u. is the reduced mass of the H$_2$+H$^-$ system. We took $J=0$ and, correspondingly, $J'=1$.  Plugging these values into Eq. (\ref{eq:cs}), we obtain the value of $10^{-9}$ a.u. for the estimated RA cross-section. The rate coefficient is estimated as $\sigma_q\cdot k/m\sim6\times 10^{-13}$ a.u. or $4\times 10^{-21}$ cm$^3/$s. In the estimation, we have neglected all the Feshbach resonances present in the collisional spectrum of H$^-$ and H$_2$. Such resonances should increase the total cross-section and will be accounted in an accurate calculation in a separate publication. 

As a summary, the main results of the article are the following. (1) We have proposed a realistic scheme for the formation of the H$_3^-$ ion in cold interstellar clouds. According to our results (2) the ion can be detected in the absorption spectrum in the millimeter wavelength range and, therefore, (3) can serve as a probe for the presence of H$^-$ in the ISM. To estimate the rate of H$_3^-$ formation in H$_2$+H$^-$ collisions,  (4) we have developed a theoretical approach to calculate the cross-section for the radiative association and determined exact and approximate quantum numbers that can be used to characterize the bound and resonant states of H$_3^-$. The developed theory is general and can be used to study other collision processes between a dimer and an atom, for instance, photoassociation in ultra-cold gases. More details about the structure of H$_3^-$ and its isotopologues, including  bound and resonant states will be presented in a separate publication, which will be relevant for considering new laboratory experiments with H$_3^-$, for example, in ion traps \cite{wester09}.

{\bf Acknowledgments.} We thank  Roland Wester for motivating us to study H$_3^-$ structure and dynamics. The study was supported by the  {\it R\'eseau th\'ematique de recherches avanc\'ees "Triangle de la Physique"}, the National Science Foundation under grant PHY-0855622, and by the programme EUROQUAM of the European Science Foundation. R.G. acknowledges  support from {\it Insitut Francilien de recherches sur les atomes froids (IFRAF)}.


\end{document}